\documentclass[aps,prd,preprint,amsmath,amssymb]{revtex4}

\def\D{D_{abc}}

\begin{document}

\title{Metric Redefinitions in Einstein--\AE ther
Theory}
\author{Brendan Z. Foster}
\email[]{bzf@physics.umd.edu}
\affiliation{Department of Physics,
University of Maryland, College Park, MD 20742-4111}

\date{July 1, 2005}


\begin{abstract}

`Einstein--\AE ther' theory, in which gravity couples to a
dynamical, time-like, unit-norm vector field, provides a means for
studying Lorentz violation in a generally covariant setting.
Demonstrated here is the effect of a redefinition of the metric
and `\ae ther' fields in terms of the original fields and two free
parameters. The net effect is a change of the coupling constants
appearing in the action.  Using such a redefinition, one of the
coupling constants can be set to zero, simplifying studies of
solutions of the theory.

\end{abstract}

\maketitle
%
%---------------------------------------------------------------
%
\section{Introduction}

Interest has lately grown in the possibility that Lorentz symmetry
is not an exact symmetry of nature.  In particular, it may be
broken by as-yet-unknown quantum gravity effects.  In an effective
field theory description, this symmetry-breaking can be realized
by a vector field that defines the ``preferred'' frame. In the
flat space-time of the Standard Model, this field can be treated
as non-dynamical, background structure. In the context of general
relativity, diffeomorphism invariance (a symmetry distinct from
that of local Lorentz invariance) can be preserved by elevating
this field to a dynamical quantity.

These considerations motivate the investigation of
``vector-tensor'' theories of gravity. One such model couples
gravity to a vector field that is constrained to be everywhere
timelike and of unit norm.  Theories providing a preferred
timelike direction are of most interest as rotational symmetry
appears to be soundly preserved.  The unit-norm condition embodies
the notion that the theory assigns no physical importance to the
norm of the vector. For a review of recent investigations of this
model, see~\cite{Eling:2004dk}. Following these authors, I shall
refer to this theory as `Einstein--\AE ther' theory, or `\AE
-theory'.

The purpose of this work is to demonstrate the effect of a field
redefinition on the conventional, second-order \AE -theory action.
The redefinition considered is of the form $g_{ab} \rightarrow
g'_{ab} = A(g_{ab} - (1-B)u_a u_b)$, $u^a \rightarrow u'^a=
(1/\sqrt{AB})u^a$, where $g_{ab}$ is a Lorentzian metric and $u^a$
is the `\ae ther' field. The action has the most general form that
is generally covariant, second order in derivatives, and
consistent with the unit-norm constraint. The redefinition
preserves this most-general form, since it preserves covariance of
the action, does not introduce higher derivatives, and preserves
the unit-norm constraint.  The net effect is then a transformation
of the coupling constants in the action.  The study of \AE-theory
systems can be simplified in certain cases by invoking this
transformation to give the couplings more convenient values;
e.g.~by setting one of the constants to zero.

This work generalizes a result of Barbero and
Villase\~{n}or~\cite{BarberoG.:2003qm} that shows equivalence
between vacuum general relativity and an \AE-theory system whose
coupling constants satisfy certain relations.  The four constants
must be specific functions of one free parameter for their result
to apply.  I consider here the general case in which the
parameters have arbitrary values. This work also uses a simpler
parametrization of the redefinition than that
of~\cite{BarberoG.:2003qm} and works with a now-more-common form
of \AE-theory action. The translation between this work
and~\cite{BarberoG.:2003qm} will be given below.
%
%-------------------------------------------------------------------
%
\section{Transformation of the Action}

The conventional, second-order \AE-theory action $S$ has the form
\begin{equation}
\label{ACT}
    S = \frac{-1}{16\pi G}\int \sqrt{|g|}\, \mathcal{L},
\end{equation}
with Lagrangian $\mathcal{L}$
\begin{equation} \label{LAG}
\begin{split}
    \mathcal{L} = R
    + &c_1 (\nabla_a u_b) (\nabla^a u^b)
    + c_2 (\nabla_a u^a)( \nabla_b u^b)\\ &
    + c_3 (\nabla_a u^b) (\nabla_b u^a)
    + c_4 (u^a\nabla_a u^c) (u^b \nabla_b u_c),
\end{split}
\end{equation}
where $R$ is the scalar curvature of the metric $g_{ab}$ (with
signature $({+}{-}{-}{-})$\footnote{The conventions
of~\cite{WALDBOOK} are followed in all other respects, including
the use of units in which c=1.}), and the $c_i$ are dimensionless
constants. This action has the most general form that is
covariant, second-order in derivatives, and consistent with the
`unit-constraint' $u^a u_a = 1$.

We will assume that the fields are on-shell with respect to this
constraint, rather than incorporate it via a Lagrange multiplier.
This approach is justified if we view two actions as equivalent if
they lead to the same equations of motion. We obtain the same
equations of motion either by subjecting the off-shell action with
a multiplier term to general variations, then solving for the
multiplier in terms of the other fields, or by subjecting the
on-shell action to variations that preserve the constraint. It
follows that two actions are equivalent if they agree on-shell.
The redefinition given below preserves the constraint; hence, it
preserves this sense of equivalence.

We begin by considering `unprimed' variables---a Lorentzian metric
$g_{ab}$ and a timelike vector field $u^a$, satisfying $g_{ab} u^a
u^b = 1$.  We then define `primed' fields:
\begin{equation}
\label{REDEF}
\begin{split}
    g'_{ab} &= A\big(g_{ab} - (1 - B)u_a u_b\big)\\
                u'^a &= \frac{1}{\sqrt{AB}} u^a
\end{split}
\end{equation}
where $A$ and $B$ are positive constants.  The sign of $A$ merely
changes the signature convention of the metric so is irrelevant. A
negative value of $B$ results in a primed metric of Euclidean
signature.  We restrict to positive $B$ to ensure comparison of
Lorentzian theories. The primed inverse-metric $g'^{ab}$ and the
primed \ae ther one-form $u'_a \equiv g'_{ab} u'^b$ are then
uniquely determined in terms of unprimed fields:
\begin{equation}\label{RED2}
\begin{split}
    g'^{ab} &= \frac{1}{A}\Bigl(g^{ab}
        - \Bigl(1 - \frac{1}{B}\Bigr) u^a u^b\Bigr)\\
                u'_a &= \sqrt{AB}\;u_a.
\end{split}
\end{equation}
It follows that $u'^a u'_a = 1$.

To observe the effect of this redefinition on the
action~\eqref{ACT}, we shall start with the primed action and
express it in terms of unprimed variables.  We shall find that the
form of the action is left invariant, with new parameters $G$,
$c_i$ given as functions of $A,\,B$, and the original $G',c'_i$.
The calculation is straightforward but lengthy---the demonstration
will be explicit to ease the checking of the final results.

Let us begin by considering the role of the parameter $A$, whose
net effect is a re-scaling of the action.  This occurs because $A$
re-scales the field variables in such a way that each term in the
Lagrangian~\eqref{LAG} acquires the same factor. Writing the
Lagrangian in terms of primed variables, then invoking the
substitutions~\eqref{REDEF} and~\eqref{RED2}, one finds that each
term in the un-primed Lagrangian carries an over-all factor of
$1/A$.  The ratio of primed-to-unprimed metric determinants will
equal $A^4$, times a $B$-dependent factor given below.  Thus, the
un-primed action~\eqref{ACT} will carry a net factor of $A$ and
will have no other $A$-dependence.  This factor can be absorbed
into a redefinition of $G$.  Having thus accounted for the effect
of $A$, we will set $A=1$ in the calculations that follow.

We can deduce the full relation between metric determinants $g$,
$g'$ by evaluating them in a basis, orthonormal with respect to
$g_{ab}$, of which $u^a$ is a member.  In this basis, $g = -1$.
From the expression $g_{ab} = u_a u_b + h_{ab}$, with $h_{ab} u^a
= 0$, we have $g'_{ab} = u'_a u'_b + h_{ab}$.  It follows that $g'
= -(u^a u'_a)^2 = -B$ in this basis. Generalizing to an arbitrary
basis, we conclude that
\begin{equation}
    g' = B g.
\end{equation}
The action then re-scales: $S' = \sqrt{B} S$. The above rescalings
effect a redefinition of Newton's constant:
\begin{equation}
    G = \frac{G'}{A \sqrt{B}},
\end{equation}
(restoring $A$ temporarily).
%
%--------------------------------------------------------------
%
\subsection{Curvature Term}

We turn now to the curvature term in the Lagrangian~\eqref{LAG}.
We start by examining properties of the redefined connection
coefficients $\Gamma^a_{bc}$,
\begin{equation}
    (\Gamma^a_{bc})' = \Gamma^a_{bc} + g^{ad} D_{dbc},
\end{equation}
where
\begin{equation}
    \D = \frac{B - 1}{2} \big(\delta^d_a -(1-1/B)
    u^d u_a\big)\big[ \nabla_b (u_a u_c) + \nabla_c (u_a u_b) -
    \nabla_a (u_b u_c) \big].
\end{equation}
Let us define the following quantities:
\begin{equation}
\begin{split}
    S_{ab} &= \nabla_a u_b + \nabla_b u_a\\
    F_{ab} &= \nabla_a u_b - \nabla_b u_a\\
    \dot u^a &= u^b \nabla_b u^a.
\end{split}
\end{equation}
We can organize $\D$ as follows:
\begin{equation}
    \D = \frac{B - 1}{2} \big( u_a X_{bc} + u_b F_{ca} + u_c
    F_{ba}\big),
\end{equation} where
\begin{equation}
    X_{bc} = \frac{1}{B} \bigl(S_{bc} +
        (B - 1)(\dot u_b u_c + u_b \dot u_c)
            \bigr),
\end{equation}
and the unit-constraint has been enforced.

We will now note some useful relations involving $\D$.  To begin,
we find that
\begin{equation}
    u^a S_{ab} = u^a X_{ab} = u^a F_{ab} = \dot u_b.
\end{equation}
We then find that contraction once with $u^a$ gives
\begin{equation}
\begin{split}
    u^a \D &= \frac{(B -1)}{2B}\big(S_{bc} -
        (\dot u_b u_c + u_b \dot u_c)\big),\\
    u^c \D &= \frac{B - 1}{2}\big(F_{ba}
            + (\dot u_a u_b + u_a \dot u_b)\big),
\end{split}
\end{equation}
and that contraction twice gives
\begin{equation}
\begin{split}\label{uuD}
    u^b u^c \D &= (B -1) \dot u_a,\\
    u^a u^b \D &= 0.
\end{split}
\end{equation}
In addition,
\begin{equation}
\begin{split}
     X^{ab}u^c \D = (B -1) \dot{u}_a \dot{u}^a, \\
      F^{ab}u^c \D = \frac{(1-B)}{2} F_{ab} F^{ab}.
\end{split}
\end{equation}
As for the trace of $\D$, we find that
\begin{equation}
\begin{split} \label{TRACE}
%    D_{ab}^{\phantom{aa}b} & \equiv g^{bc}\D =
%    \frac{B-1}{B}\bigl(u_a \nabla_b u^b + B \dot u_a\bigr)\\
    D^b_{\phantom{a}bc} &\equiv g^{ab} \D = 0.
\end{split}
\end{equation}

Let us now examine the transformation of the curvature tensor.  A
short calculation reveals that
\begin{equation}
    (R_{abc}^{\phantom{abc}d})' = R_{abc}^{\phantom{aaa}d}
            + 2 \nabla^{\phantom{a}}_{[b} D^d_{\phantom{a}a]c}
            + 2 D^d_{\phantom{a}e[b} D^e_{\phantom{a}a]d},
\end{equation}
so that
\begin{equation}
    (R_{ab})' = R_{ab} + W_{ab},
\end{equation}
where
\begin{equation}
   W_{ab} = \nabla_d D^d_{\phantom{a}ab} -
            D^d_{\phantom{a}e a}D^e_{\phantom{a}d b}.
\end{equation}
The scalar curvature $R' = R'_{ab}\, g'^{ab}$ takes the form
\begin{equation}
\label{SCR}
    R' = R_{ab}g^{ab} + \frac{1-B}{B} R_{ab} u^a u^b
        + W_{ab}
        \left( g^{ab} + \frac{1-B}{B} u^a u^b\right).
\end{equation}
The second term on the right-hand-side can be re-expressed via the
definition of the curvature tensor:
\begin{equation}\begin{split}
    R_{ab} u^a u^b &= u^a \nabla_b \nabla_a u^b
                - u^a \nabla_a\nabla_b u^b \\
                & = (\nabla_a u^a) (\nabla_b u^b)
                    - (\nabla^a u^b)( \nabla_b u_a) + \upsilon,
\end{split}
\end{equation}
where $\upsilon$ represents a total divergence.  We can discard
this, with the same justification given above for taking the
fields as on-shell. The symbol $\upsilon$ will continue to
represent other total divergences that appear in the calculations
below, but the specific form of the divergence will differ by
equation. The third term on the right-hand-side of~\eqref{SCR} has
the form
\begin{equation}
\begin{split}
    W_{ab} g^{ab} & = -D^{cba}\D + \upsilon \\
            & = \frac{(1-B)}{2}\left(u^c X^{ab}
                + u^bF^{ac}\right)\D + \upsilon \\
        &=\frac{-(1-B)^2}{2}\left(\dot u^a \dot u_a
                - \frac{1}{2} F_{ab}F^{ab}\right) + \upsilon.
\end{split}
\end{equation}
As for the last term in~\eqref{SCR}, we have
\begin{equation}
\begin{split}
    W_{ab} u^a u^b & =u^a u^b (\nabla^c D_{cab}
                 - D_{cda}D^{dc}_{\phantom{aa}b})\\
        &= -\D u^c (2 \nabla^a u^b + D^{bad} u_d) + \upsilon\\
         &= \frac{1-B}{2} \bigl((\dot u_a u_b + u_a \dot u_b)+
            F_{ba}\bigr)\\
                    &\qquad \times\bigl( S^{ab} + \frac{B -1}{2}
                    (\dot u^a u^b + u^a\dot u^b)
                + (\frac{B+1}{2})F^{ab}\bigr) + \upsilon\\
        &= \frac{(B^2 -1)}{2} \bigl( \dot u^a \dot u_a -
                \frac{1}{2} F_{ab} F^{ab} \bigr) + \upsilon.
\end{split}
\end{equation}
Combining the above and suppressing a total divergence, we can
express the transformation of the scalar curvature as
\begin{equation}
\label{SCR2}
\begin{split}
    R' &= R + \frac{1-B}{B}\bigl((\nabla_a u^a) (\nabla_b u^b) - (\nabla^a u^b)( \nabla_b u_a) \bigr)
           + \frac{(1-B)^2}{2B}\bigl(\dot u^a \dot u_a  -\frac{1}{2}F_{ab}F^{ab}\bigr)\\
        &= R - \frac{1-B}{2B}\Bigl\{(1-B)(\nabla_a u_b) (\nabla^a u^b)
            - 2(\nabla_a u^a)(\nabla_b u^b)\\
            &\qquad\qquad\qquad\qquad +(1+B)(\nabla^a u^b)(\nabla_b u_a)
            -(1-B)(\dot u^a \dot u_a)\Bigr\}.
\end{split}
\end{equation}

We can extract from this expression contributions $a_i$ to the
redefined $c_i$:
\begin{equation}\label{AAA}
\begin{split}
    a_1 &= -\frac{(1-B)^2}{2B}\\
    a_2 &= \frac{1-B}{B}\\
    a_3 &= -\frac{1-B^2}{2B}\\
    a_4 &= \frac{ (1-B)^2}{2B}.
\end{split}
\end{equation}
The constants $a_i$ are characterized by the relations
\begin{equation}
\label{GREQ}
 0=a_1 + a_4 = a_1 + a_2 + a_3 = a_1(a_1 - 2) -
(a_3)^2,
\end{equation}
and $a_1 < 0$.  If the $c_1$ satisfy these conditions, then the
\AE -system is equivalent to pure gravity via a field
redefinition.  The translation from this result to that
of~\cite{BarberoG.:2003qm} is made by choosing $A =
-\sqrt{|\alpha(\alpha+\beta)|}/2$ and $B = -\alpha/(\alpha +
2\beta)$ [compare the first line of~\eqref{SCR2} with Eqn.~(6)
of~\cite{BarberoG.:2003qm}].
%
%----------------------------------------------------------
%
\subsection{\AE ther Term}

We now proceed to examine the transformation of the \ae ther
portion of the Lagrangian.   From the form of the covariant
derivative
\begin{equation}
    (\nabla_a u^b)' = \frac{1}{\sqrt{B}} \big(\nabla_a u^b
            + D^b_{\phantom{a}ac} u^c\big),
\end{equation}
and the relations~\eqref{TRACE} and~\eqref{uuD}, we can deduce the
transformation of the $c_2$ and $c_4$ terms:
$(\nabla_a u^a)' = (1/\sqrt{B})(\nabla_a u^a)$,
$(\dot u^a)' = \dot u^a$
and further
$(\dot u_a)' = \dot u_a$.
Thus, we have
\begin{equation}
    \big((\nabla_a u^a) (\nabla_b u^b)\big)'
    = \frac{1}{B} \big((\nabla_a u^a) (\nabla_b u^b)\big),
\end{equation}
and
\begin{equation}
    \big(\dot u^a \dot u_a\big)'
    = \big(\dot u^a \dot u_a\big).
\end{equation}
These results indicate contributions of $c'_2/B$ to $c_2$ and
$c'_4$ to $c_4$.

It will be convenient to reorganize the $c_1$ and $c_3$ terms:
\begin{equation}
    c_1 (\nabla_a u_b)( \nabla^a u^b) + c_3(\nabla_a u_b )(\nabla^b u^a) =
    \frac{c_+}{4} S_{ab}S^{ab} + \frac{c_-}{4} F_{ab} F^{ab},
\end{equation}
where
$c_{\pm} = c_1 \pm c_3$.
We then need the form of the covariant derivative of $u'_a$,
\begin{equation}
\begin{split}
    (\nabla_a u_b)' &= \sqrt{B} \big(\nabla_a u_b
            - D_{cab} u^c\big)\\
            & = \frac{1}{2\sqrt{B}}\bigl(S_{ab}
                    + (B -1)(\dot u_a u_b + u_a \dot u_b)\bigr)
                   + \frac{\sqrt{B}}{2} F_{ab}.
\end{split}
\end{equation}
Raising an index on the symmetrized derivative,
\begin{equation}
    (S_{cb} g^{ac})' = \frac{1}{\sqrt{B}} \big(S^a_{\phantom{a}b}
            + (B -1) u_b \dot u^a\big),
\end{equation}
leads to
\begin{equation}
    (S_{ab} S^{ab})' = (S^a_{\phantom{a}b} S^b_{\phantom{a}a})' =
        \frac{1}{B} \left(S_{ab} S^{ab}
        + 2 (B -1) \dot u^a \dot u_a\right),
\end{equation}
indicating contributions of $c'_+/B$ to $c_+$ and $(B-2)c'_+/2B$
to $c_4$.  Raising an index on the anti-symmetrized derivative,
\begin{equation}
    (F_{cb} g^{ac})' = \sqrt{B}
        \left(F^a_{\phantom{a}b} +
        \frac{1-B}{B} u^a \dot u_b \right),
\end{equation}
leads to
\begin{equation}
    (F_{ab} F^{ab})' = -(F^a_{\phantom{a}b} F^{b}_{\phantom{b}a})'
        = B \left(F_{ab} F^{ab}
        + 2\frac{1-B}{B} \dot u^a \dot u_a\right),
\end{equation}
indicating contributions of $Bc'_-$ to $c_-$ and $(1-B)c'_-/2$ to
$c_4$.

Collecting the above results, we find contributions $b_i$ to the
redefined $c_i$:
\begin{equation}
\label{BBB}
\begin{split}
    b_1 & = \frac{1}{2B}\bigl(c'_+ + B^2 c'_-\bigr)\\
                & = \frac{1}{2B}\bigl((1+B^2)c'_1
                    + (1-B^2)c'_3\bigr)
\\
    b_2 &= \frac{c'_2}{B}\\
    b_3 & = \frac{1}{2B}\bigl(c'_+ - B^2 c'_-\bigr)\\
                & = \frac{1}{2B}\bigl((1-B^2)c'_1
                    + (1+B^2)c'_3\bigr)
\\
    b_4 &= c'_4 - \frac{1-B}{2B}\bigl(c'_+ - B c'_-\bigr)\\
        & = c'_4 - \frac{1 - B}{2B}\bigl((1-B) c'_1
            +(1+B) c'_3\bigr).
\end{split}
\end{equation}
The redefined $c_i$ are given by the sum of $a_i$~\eqref{AAA} and
$b_i$~\eqref{BBB}:
\begin{subequations}
\label{CCC}
\begin{equation}
\begin{split}
    c_1 &=\frac{1}{2B}\bigl(c'_+ + B^2 c'_-
                - (1-B)^2\bigr)\\
    &=\frac{1}{2B}\big((1+B^2)c'_1
                    + (1-B^2)c'_3 - (1-B)^2\big)\\
    c_2 &=\frac{1}{B}\big(c'_2 + 1 - B\big)\\
    c_3 &= \frac{1}{2B}\bigl(c'_+ - B^2 c'_-
                -(1-B^2)\bigr)\\
           &=\frac{1}{2B}\big((1-B^2)c'_1
                    + (1+B^2)c'_3 - (1-B^2)\big)\\
    c_4 &= c'_4 - \frac{1-B}{2B}\bigl( c'_+ - Bc'_-
                -(1-B)\bigr)\\
        &=   c'_4 - \frac{1}{2B}\big((1-B)^2 c'_1
            +(1-B^2) c'_3 - (1-B)^2\big).\\
\end{split}
\end{equation}
In addition,
\begin{equation}
\begin{split}
    c_{+} &= \frac{1}{B}\bigl(c'_{+} - (1-B)\bigr)\\
    c_- & = B c'_- + (1-B)
\end{split}
\end{equation}
\end{subequations}
%
%---------------------------------------------------------------
%
\section{\label{DISC}Discussion}

The redefinition~\eqref{REDEF} can simplify the problem of
characterizing solutions for a specific set of $c_i$. This is done
by transforming that set into one in which the $c_i$ take on more
convenient values.  It was noted in~\cite{BarberoG.:2003qm} that a
system with restricted values of the coefficients, equivalent to
$c_i$ that satisfy~\eqref{GREQ}, can be transformed into \ae
ther-free general relativity. The current work extends this result
by allowing for general values of the $c_i$.  Using this result,
different sets of $c_i$ are seen to be equivalent. For example, it
follows from the relations~\eqref{CCC} that a set of $c_i$ is
equivalent to one in which one of $c_+$, $c_-, \text{ or } c_2$
vanishes if the original values satisfy, respectively, $c_+ < 1,
c_- < 1, \text{ or } c_2 > -1$.

An extra constant can be eliminated in the case of
spherically-symmetric configurations~\cite{Eling:2003rd}. In this
case, the hyper-surface orthogonality and unit norm of the \ae
ther imply the vanishing of the twist $\omega_a =
\epsilon_{abcd}u^b \nabla^c u^d$, so that
\begin{equation}\label{TWIST}
    \omega_a \omega^a = \dot u^a \dot u_a - 1/2 F^{ab} F_{ab} = 0.
\end{equation}
The redefinition of a particular configuration preserves any
Killing symmetries shared by the metric and \ae ther fields, so it
preserves the relation~\eqref{TWIST}.  One can then eliminate, for
instance, $c_+$ by redefinition and $c_4$ by absorption into
$c_-$.  The Lagrangian is reduced to the form
\begin{equation}
    {\mathcal L} = R + \frac{c_-}{4} F_{ab}F^{ab} + c_2 (\nabla_a
    u^a)^2.
\end{equation}
This is considerably simpler than the general form~\eqref{LAG},
since the connection enters the \ae ther action only through the
divergence of $u^a$.

Spherically-symmetric, static \AE-theory black hole solutions have
been shown to exist~\cite{Eling:2004dk}.  One can apply the above
to simplify their study.  The question arises, though, of how to
define the location of the horizon.  Initial results~\cite{Chris}
indicate that a solution with a horizon can be equivalent to one
without, where the horizon is defined via the fastest speed of
linearized wave modes.

Once non-\ae ther matter is included, a metric redefinition not
only changes the $c_i$ coefficients, but also modifies the matter
action. The fact that Lorentz violating effects in
non-gravitational physics are already highly
constrained~\cite{LOR} means that, to a very good approximation,
there is a universal metric to which matter couples. Within the
validity of this approximation, one can identify the field
$g_{ab}$ with this universal metric, thus excluding any \ae ther
dependence from the matter action. This identification then
eliminates the freedom to redefine the metric.  Recent studies of
observational bounds on the values of the $c_i$, such
as~\cite{TEST}, have adopted this convention.
%
%--------------------------------------------------------------------
%
\begin{acknowledgments}
I would like to thank Ted Jacobson for discussions and impeccable
editing advice.
\end{acknowledgments}
%
%-------------------------------------------------------------------
%

\end{document}